\documentclass[12pt,preprint]{aastex}

\shorttitle{Dust Shell of IRAS 02091+6333}
\shortauthors{Zijlstra \& Weinberger}

\def\smy{{\,\rm M_\odot yr^{-1} }}

\begin{document}

\title{A wall of dust around  a proto-Mira}
\author{Albert A. Zijlstra}
\affil{Astrophysics Group, Department of Physics, UMIST, P.O. Box 88,
 Manchester M60 1QD, UK}
\email{a.zijlstra@umist.ac.uk}

\and

\author{R. Weinberger}
  \affil{Institut f{\"u}r Astrophysik, Leopold-Franzens-Universit{\"a}t
Innsbruck, Technikerstrasse 25, A-6020 Innsbruck, Austria}
\email{ronald.weinberger@uibk.ac.at}

\begin{abstract}

 We present the discovery of a huge (19$\arcmin$ $\times$ 16$\arcmin$)
dust ring surrounding a bright ($V$ = 10.60) red star.  The dust ring
has, at $D = 700\,$pc, a diameter of 4 pc, and a central hole of
$\sim$1.5 pc across.  Part of the shell is also seen as an absorption
nebulosity.  The star is classified as a M3III AGB star. Among AGB
stars its detached shell is of unrivalled size.  Detached shells
around AGB stars are normally interpreted in terms of thermal
pulses. However, in this case a significant fraction of the shell may
consist of swept-up ISM; the detached appearance can be explained with
wind--ISM interaction. We present a model where the AGB wind has been
stopped by the surrounding ISM, and the swept-up shell is now
expanding at the sound speed.  The model predicts that the ring will
disperse over a few times $10^5\,$yr, and eventually will leave
a large hole in the ISM surrounding the AGB star or its
future planetary nebula.

\end{abstract}


 \section{AGB Supershells}

 Mass loss dominates stellar evolution on the Asymptotic Giant Branch
(AGB).  The mass loss probably begins on the early AGB, or even the
first Red Giant Branch \citep{J99}, but culminates during the last
10\%\ of the AGB, the thermal-pulsing AGB (TP-AGB).  Observed
mass-loss rates range from $10^{-7} \smy$ for semiregulars and Miras,
up to a few times $10^{-4} \smy$ for OH/IR stars and obscured carbon
stars. Typically, the Mira phase lasts between $7 \times 10^4\,\rm yr$
and a few times $10^5$ yr (e.g. \citealt{Woo90}). The OH/IR star or
the obscured carbon star phase may last $\sim 10^4\,$yr.  Because the
mass-loss rates increase during the AGB evolution, circumstellar
shells may be expected to have a centrally-condensed, compact
appearance. But mass-loss fluctuations during the thermal-pulse cycle
can lead to detached circumstellar shells \citep{VW93}, and
observations of detached shells have been interpreted as such
\citep{ZLWJ92}.

Evidence for detached shells came from the IRAS point source
data, when \citet{WJ86} found carbon stars with excess flux at 60$\mu$m.
\citet{ZLWJ92}, in the same way, found  oxygen-rich
stars (M stars) with detached shells; they interpreted these in terms
of mass-loss variations during the thermal-pulse cycle.  From IRAS
spectroscopy \citet{HIKB98} found evidence for a detached shell with a
small inner radius around the O-rich star R Hya (see also
\citealt{ZB02}).

Photometric evidence may be affected by cirrus \citep{IE95}, and
direct imaging can give more stringent evidence for detached shells.
Izumiura et al. (1996, 1997) imaged detached shells around the carbon
stars Y CVn and U Ant, with U Ant even showing a {\it double} detached
shell.  A spectacular thin CO ring was found around the carbon star TT
Cyg \citep{OBL98,OBL00}. \citet{LOL99} found a detached shell around
the carbon star U Cam.  \citet{SMK00} reported parsec-sized thin dust
shells around two carbon-rich post-AGB stars, AFGL 2688 and AFGL 618,
indicating episodic mass-loss variations with timescales of a few
times 10$^4$ years. A thermal-pulse explanation is likely for all
these cases.

Circumstellar shells could grow to very large dimensions during the
AGB evolution. In principle, a gradual and gentle slow-down of the
earliest ejected shell(s) by the ISM could allow subsequently expelled
matter to catch up, and yield very large, detached-appearing shells
independent of the thermal-pulse cycle. \citet{YPK93}, using IRAS
survey data for 512 AGB stars and young PNe (mostly within 1 kpc),
found a total of 76 stars with circumstellar shells greater than
2$\arcmin$ in the 60$\mu$m data. Of these, four shells are larger
than 4 pc in diameter; all four appear detached, having inner
diameters of between 0.5 and 1.4 pc. (The detection of these shells is
complicated by the often very bright central sources, requiring
deconvolution techniques to separate the shell from the PSF of the
stellar source).

Such fossil shells carry the imprint of the mass-loss history and the
structure and density of the ISM.  They could help to explain the
presence of halos around planetary nebulae (but see \citealt{CSS00})
and the interaction of the most evolved of them with interstellar
matter. Clearly, a search for extended shells - the larger, the better
- is of interest.

Here we report on an extreme dust shell around the source
IRAS~02091+6333, with a diameter of 4 pc.  We interpret the shell not
in terms of mass-loss variations, but as a wall of swept-up ISM. We
suggest that rings and walls around evolved stars may not have a
single explanation.

 \section{Observations}

\subsection{The Dust Shell}

The dust ring was discovered by one of us (RW) during a search for
extremely extended structures, such as holes and emission regions
around planetary nebulae, on the IRAS SkyView maps provided by
NASA. It is visible at 60$\mu$m and - slightly better - at 100$\mu$m,
and attracted our attention because of its well-defined shape and the
presence of a stellar, almost centrally located object, prominent at
12$\mu$m but fainter at 25$\mu$m and absent at longer wavelengths. The
shape of the dust emission suggests that the central source is
responsible for the ring, which has the appearance of a detached
shell. The shell is brightest at 100$\mu$m indicative of cold
dust. Interestingly, the shell, particularly its brightest part
towards the south, is also optically identifiable as an absorbing cloud on
the POSS\,I blue-sensitive (O) plates.  The stellar field is very
dense (consistent with low foreground extinction for a line of sight
near the Galactic plane) and the lack of blue stars in the small patch
is highly noticeable. We estimate from the appearance of the patch
that $A_{\rm V} \sim 0.5$mag.

The processed IRAS images, retrieved from the IPAC SkyView images, and
also remade from the original (recalibrated) scans using the IRAS
Software Telescope \citep{ABJ95}, are shown in Fig. 1. The emission at
100$\mu$m has a slightly elliptical shape, with the long axis in the
N-S direction. It consists of a ring of emission which is thicker (and
brightest) in the south and least thick in the west. There is a
distinct, approximately round, central cavity. The outer dimensions
are 19$\arcmin$ $\times$ 16$\arcmin$, corresponding to 3.9 $\times$
3.2 pc at $D = 700\,$pc (see below). The ring has a mean apparent
thickness of $\sim$1.0 pc, and the cavity is $\sim$1.6 pc in
diameter. The resolution of IRAS at 100-$\mu$m is sufficient to
resolve the cavity, but the thickness of the ring is likely affected
by the IRAS beam.  The emission spur towards the south-west is
probably unrelated (it has a different colour temperature).  The total
flux of the ring is about 20 Jy at 60$\mu$m and 70 Jy at 100$\mu$m:
both values are uncertain due to the need to subtract a sloping
background.

     \begin{figure*}
     \label{iras.images}
     \caption{A 4-color image of  1$^\circ \times
     1^\circ$ wide containing the dust ring taken from NASA`s SkyView
     and processed with the Maximum Entropy Method to improve the
     resolution. The position of the red giant star is indicated. The
     object is located 2.3$^\circ$ north of the Galactic plane which
     runs roughly E-W.}
      \includegraphics[width=\textwidth]{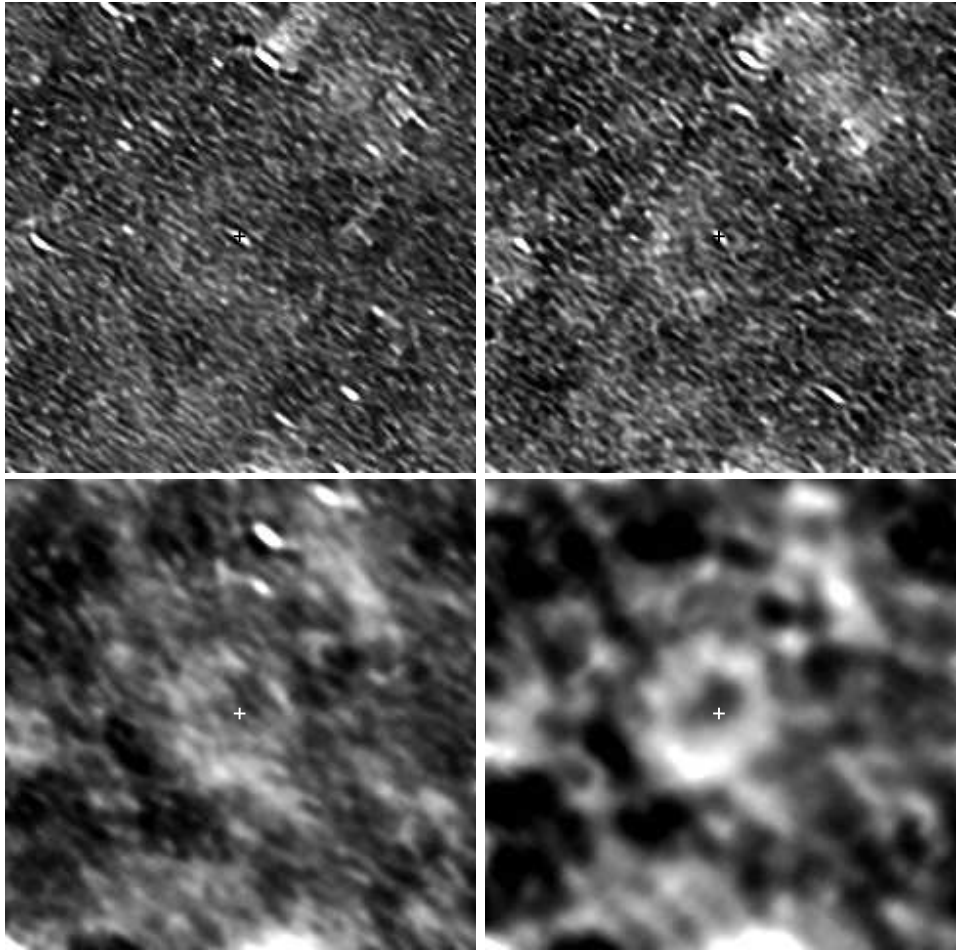}     
     \end{figure*}

The concave thickening of the ring in the south is suggestive of an
interaction with the ISM of the Galactic plane which runs
approximately E-W, 2.3$^\circ$ below the object. But this interaction
has had no marked effect on the outer shape of the ring.

A subtraction of the 60-$\mu$m from the 100-$\mu$m map, carried out
using the unprocessed images, showed that there is no obvious radial
dependence within the ring.  The black-body dust temperature is about
35 K.  The large distance from the central star indicates the shell is
heated externally (e.g., \citealt{YPK93,SMK00}). The temperature
should be compared to typical values for Galactic cirrus emission of
15--20 K \citep{B01}.

\subsection{The Central Star}

The flux densities of the central source (Table 1) are consistent with
photospheric emission. Optical images from the Palomar Sky Surveys
show a star coincident with the central source. Although rather
bright, very red, and located not far from the Galactic plane ($\ell =
131.79, b = +2.32$) in one of the most studied regions of the Milky
Way, SIMBAD does not list individual observers. It is present in the
Guide Star Catalog (GSC-id 0404101743), in the AC2000, and was covered
by the TYCHO experiment onboard of HIPPARCOS. Data for the star are
listed in Table 1.

One long-slit spectrum (300 sec) of the star was taken with the
instrument AFOSC attached to the 1.82 m telescope of the Asiago
Observatory. It was obtained on Dec 1$^{\rm st}$ 1999, at an air mass
of $\sim$1.4. The spectral range covered was 3940 -- 7860 \AA\, and
the dispersion was 4.9 \AA/pixel.  Another spectrum was kindly
provided by Maurice Gavin, taken on Dec 21st, 1999 at the Worcester
Park Observatory (UK) at a dispersion of 40 \AA/pixel. Both spectra
were reduced in the usual way.  The spectra show only the star itself,
i.e. we found no traces of any nebulosity in its immediate
neighborhood. The TiO bands are very well visible and establish the
spectral type as M3 or M4. We could however not determine the
luminosity class. The discussion below shows that a giant is most
probable and we will assume the object to be an M3\,III star. Several
photometric runs in BVRI were kindly done for us by Roger Pickard, who
used his 40 cm Newtonian Reflector. He reported $B-V$ = 2.06.

\begin{table}
\caption[]{\label{star.data} Parameters for the central star}
\begin{flushleft}
\begin{tabular}{llllll}
\hline
Position    &   $\alpha = 02^{\rm h}~ 12^{\rm m}~ 58.43^{\rm s}$
                                    &  $l = 131.79^{\rm o} $ \\
 ~~(J2000)  &    $\delta = +63^{\rm o}~ 47^\prime~ 02.5^{\prime\prime}$
                                    &  $b = +2.32^{\rm o}$ \\
proper motion         & \multicolumn{2}{l}{
                   $\Delta \alpha = 13.3 \pm 3.7\rm \, mas\, yr^{-1}$} \\
                      & \multicolumn{2}{l}{
                   $\Delta \delta = -6.6 \pm 3.6\rm \, mas\, yr^{-1}$} \\
Photometry     &   Tycho      &    GSC           &   Pickard \\
             B &  $ 12.73 \pm 0.23 $ &           &  12.55/11.7 \\
             V &  $ 10.60 \pm 0.05 $ & 9.96/10.12   & 10.5--10.1 \\
             R &                      &             &   9.5/9.3 \\
             I &                      &              &  7.4/7.3 \\
IRAS           12$\mu$m &   1.28 Jy  &  MSX  8.28$\mu$m &   1.77 Jy  \\
\phantom{IRAS} 25$\mu$m &   0.36 Jy  & \phantom{MSX} 14.65$\mu$m
& 1.24 Jy\\
E(B-V)        &  0.42 \\
Spectral type  &  M3III \\
distance      &  700: pc \\
\hline
\end{tabular}
\end{flushleft}
\end{table}

The central star appears to have a slight offset of $\sim 1.5^\prime$
from the center of the shell. However, this apparent offset is in part
caused by the difference in brightness between the N and S part of the
shell.  Compared to the peak brightness positions, the star is very
well centred, 7$^\prime$ from the two components.

\subsection{Association or Confusion?}

Although the star is close to the centre of the ring, in such a dense
stellar field the chance of an accidental superposition of the shell
and the central source has to be addressed. The IRAS catalogue lists
49 sources located within $1^{\rm o}$ radius of the present position.
Of these, 36 are detected at 12$\mu$m and 20 are detected at both 12
and 25$\mu$m. One of these has the colours of an HII region and can be
removed. Of the 36 candidates, 14 appear to be associated with a GSC
star of $m_{\rm V}<12$, within 5 arcsec. The V--[12] colours indicate
that 6 of these are early-type stars. The remaining 8 stars (including
IRAS~02091+6333) have the colours of RGB or AGB stars: the
density of such stars in this field is 1 per 1400 square
arcminutes (we note that the association of IRAS~02091+6333 is,
at 1.6 arcsec, the closest of the 14 sources.)

Using the apparent offset of 1.5$^\prime$, the chance that the central
star and ring are unrelated is 0.5\%: the association is significant
at the 3$\sigma$ level. Even without the presence of an optical star,
the significance of an association between the ring and the central
IRAS source would be close to 3$\sigma$. A relation between the ring
and its bright central AGB or RGB star is therefore plausible.

One should however also take into account how common such rings are in
the IRAS images. This close to the Galactic plane, the background
emission is extensive and structured. But within a field of $2^{\rm o} 
\times 2^{\rm o}$, there is no other ring-like structure: the background
emission shows elongated ridge-like structures instead. The difference
between these structures and the ring is shown by the fact that most
of the IRAS PSC sources in this region correspond to parts of the
ridges, tracing brightness fluctuations which are common features of
these structures. In contrast, the ring is quite smooth and was not
picked up as one or more point sources.

Although confusion can not be fully ruled out, a physical association
between star and ring appears plausible.

\section{Results}
\subsection{The Star}

For an M4III or M3III star, $ \rm (B-V)_0 = 1.75 $ or $1.71$
respectively \citep{WCV92}.  The observed $ \rm B-V = 2.134$, gives
an extinction $E(B - V) = 0.42$ or $A_{\rm V} = 1.30$, for an M3III
star. The dereddened V$-[12]$ colour of 4.9 indicates an M3 rather
than M4 star, with $M_{\rm V} = -0.6$ (e.g \citealt{WCV92}) for which we
find a distance of $D \sim700\,$pc. This extinction and distance are
not unreasonable if compared to nearby stars of known spectral type,
luminosity class and with reliable photometry: within a radius of
40$\arcmin$ four such stars were found in SIMBAD: two are about $500\,$pc
distant and have an $E({\rm B - V})$ of 0.15 and 0.30, respectively,
and the other two are about 1 kpc away and have values of 0.59 and
0.43, respectively.  Some of the extinction towards the central star
may be circumstellar. We note that the W3--5 complex, located only 2
degrees away in the Perseus arm, is at a distance of 2.35 kpc
(e.g. \citealt{HT98}).  If an extinction sheet is associated with this
spiral arm, IRAS~02091+6333 is likely located in front of it. The OB
association Cam OB1 is found a few degrees away at 900 pc \citep{Z99}.

The likely distance suggests that the M3 star is of luminosity class
III rather than I.  Furthermore, its circumstellar obscuration cannot
be very large.  The IRAS PSC flux densities and the V--[12] colour are
consistent with photospheric emission from an M3 star: there is no
evidence for hot circumstellar dust, and the present mass-loss rate
must be small. The MSX flux densities (Table 1) agree with IRAS but
the 14 micron point seems a little higher, a difference that may be
caused by the larger beam width of MSX, which was a smaller telescope.

Evidence for mass loss (the detached shell), generally indicates a
thermal-pulsing, variable AGB star. \citet{J99} shows that some RGB
stars also show dust emission, but only close to the star.  But there
is no strong evidence for variability of the M3III star. The 228 TYCHO
measurements include some fainter ones, but the magnitude is close to
the limiting magnitude of the instrument and the catalog does not flag
the star as variable. A comparison of its brightness in a POSS\,I
overlapping region (plates E/O 597, taken on 1952-09-15/16, and E/O
878, taken on 1953-10-30/31) also does not show any obvious
variability. Monitoring by Roger Pickard between January 2000 and
January 2001 indicates a possible brightening at V from 10.5 to 10.1,
and from 12.5 to 11.7 at B, but this requires further confirmation.

The bolometric magnitude is estimated as $M_{\rm bol} = -3.2$, based
on the BC(V) of \citet{Flo96}.  This is very uncertain as the
bolometric correction is a steep function of colour. The corresponding
luminosity $L\approx 1500\rm \, L_\odot$ is, given the uncertainty,
consistent with a TP-AGB star (e.g.  \citealt{WS90}).  Low-level
variability would be consistent with a pre-Mira evolutionary stage.
But the data does not rule out a location near the tip of the RGB;
neither can a TP-AGB star during the helium-burning phase of the TP
cycle be ruled out, when luminosity and the mass-loss rate are
reduced.

\subsection{ The Shell}

The fact that part of the shell is seen in absorption gives an
indication of its mass. We use the conversion ratio $A_{\rm V}/N({\rm
H}) = 5.3\times 10^{-22} \rm\, cm^2\, mag$. The absorption region is
roughly $3^\prime \times 3^\prime \approx 0.4\, \rm pc^2$. This gives a
total mass of $3\, M_\odot$ for $A_{\rm V}=0.5\,$mag. This estimate is
by necessity rough, since the extinction is estimated and the size
of the region has significant uncertainties.

The IRAS flux (70 Jy at 100 $\mu$m) gives a second determination of
the mass of the shell. Using a black-body dust temperature of 35 K, we
derive a total dust mass of $0.025 d^2\,\rm M_\odot$, where $d$ is the
distance in kpc \citep{T00}.  For a distance of $700\,$pc and a
gas-to-dust mass ratio of 200, the total mass in the ring is $\sim
2.5$ M$_\odot$. Most of this mass will be in the brighter southern
region. The agreement between the two estimates is deceptive,
since the uncertainties in these calculates are easily a factor of
2. However, it indicates a relation between the absorption region and
the ring.

\subsection{Shell Origin}

In general, distant shells around AGB stars have been interpreted in
terms of episodically enhanced mass loss, related to the thermal-pulse
cycle \citep{ZLWJ92,IWJ97}.  Assuming an expansion velocity
of 10 km/s, the outermost gas in the shell would have been ejected $2
\times 10^5\,$yr ago, and the gas closest to the star $1 \times
10^5\,$yr ago. The duration of the mass-loss event would not be
inconsistent with the duration of the thermal-pulse cycle
\citep{VW93}. However, this would require an equally long time of
quiescence to explain the cavity, whilst the duration of the
low-luminosity part of the cycle, during helium burning, is only
10--15 per cent of the cycle. 

The relatively large mass of the shell suggests that much of the the
gas does not originate from the star. Given the uncertainties in
distance, integrated IRAS flux density, and temperature, the mass
could be as low as 1 M$_\odot$ in which case a significant
contribution from stellar mass loss could not be excluded.

\section{The  Wall}

\subsection{AGB--ISM Interaction}

In view of these two points, we suggest that the shell reflects ISM
sweep-up. In this model, the AGB wind, moving out at $\sim 10\rm\,
km\,s^{-1}$ into the ISM, is creating a a denser shell consisting of
both components.  Models for the early phases of such sweep-up have
been calculated by \citet{YPK93}. The typical cooling time for the
post-shock region is $10^4\,\rm yr$, with density increasing to
$10^2$, reducing as the front slows down \citep{Spi78}. The
interaction region can thus be considered momentum-driven, in
agreement with \citet{YPK93}. The velocity of the interface declines
as $v \propto r^{-1} $ or $v \propto t^{-0.5}$, until its velocity
drops below the sound speed in the ISM outside the shell.
\citet{YPK93} do not consider cases where the amount of matter swept
up from the ISM exceeds the mass contributed by the stellar wind. The
large mass of the present shell suggests the present shell has evolved
past this phase.

     \begin{figure}
     \label{momentum.flow}
     \caption{Early phases of the growth of a wall around an AGB star,
      showing radius, mass and velocity of the swept-up wall.  Assumed
       parameters are: stellar mass-loss rate $5 \times
       10^{-8}\,\rm M_\odot \, yr^{-1}$, wind velocity $10 \,\rm km
       \,s^{-1}$ and ISM density $n = 4 \,\rm H \, cm^{-3}$. This phase
       ends when the velocity drops below the sound speed of the ambient 
      medium.}
      \includegraphics[width=8.5cm]{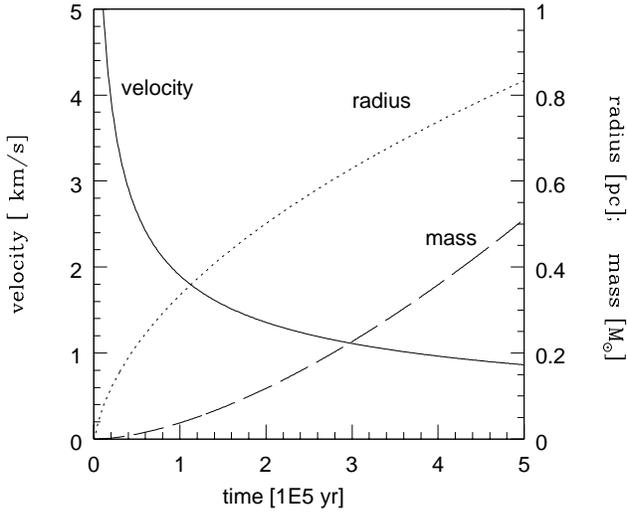}     
     \end{figure}

Using the measured diameter of the shell, and assuming its mass
reflects the ISM, suggests a local ISM density (outside of the shell)
of a few particles cm$^{-3}$. Fig. \ref{momentum.flow} shows the early
development of the wall. In this case, the velocity of the interface
(the 'wall') drops to 1\,km\,s$^{-1}$ after about $5 \times
10^5\rm\,yr$.  The early phase ends once the thermal pressure $P$ of
the ambient ISM becomes non-negligible, causing the interface to
stall.  This happens once

\begin{equation}
  P \approx \frac{ \dot M  v_{\rm w}}{4 \pi r^2} 
\end{equation}

For the ambient pressure, we use a typical value of $n T = 100$.
Assuming that the interface has stalled at a distance of 2\,pc from
the star (the present radius of the shell), the equation above yields
a mass-loss rate $\dot M \approx 5 \times 10^{-8} \rm \,
M_\odot\,yr^{-1}$. This value is typical for semi-regular variables,
which have not yet evolved into Mira variables, and is sufficiently
low that the 12-$\mu$m emission would still be dominated by the
photosphere. Much of the stellar envelope mass on the AGB may be lost
in such a moderate wind before the onset of the Mira phase (the narrow
$PL$ relation for short-period Miras \citep{GL81,FGWC89} suggests that
their masses are already reduced to close to the final white dwarf
mass before the Mira phase begins).  IRAS~02091+6333 could have built
a hydrogen wall during its pre-Mira evolutionary phase, before
reaching the mass-loss rates associated with the tip of the AGB.

However, the resulting shell will not be stable: once the interface
becomes trapped, the swept-up shell begins to expand into the ambient
ISM at its sound speed.  Using a typical value of the sound speed of
$c_{\rm s} = 1\rm \,km\,s^{-1}$, the ring reaches a thickness of 1\,pc
in $10^6\,\rm yr$. As a rough estimate, we may assume that the shell
ceases to be observable once its density has dropped to twice that of
te surrounding (presumably uniform) medium. For a thin shell:

\begin{equation}
  4 \pi r^2 c_{\rm s} 2 \rho_0 = \frac{4 \pi}{3} r^3 \rho_0 
\end{equation}

\noindent which gives

\begin{equation}
   t = \frac{1}{6} \frac{r}{c_{\rm s}} \approx 3 \times 10^5\,\rm yr
\end{equation}

The observed thickness of the ring is roughly 1 pc, but this
is affected by the IRAS 10-$\mu$m resolution and the actual thickness is
probably rather less. Within the limited accuracy, we find agreement
between the possible parameters of such a wall and the observed ring
of IRAS~02091+6333.

This suggests that it is possible to create such a wall even before
the Mira phase on the AGB, and that the wall could survive for an
appreciable fraction of the thermal-pulsing AGB ($\sim 10^6\,\rm yr$).
At a constant mass-loss rate, the interface will stall in less than
$10^6\rm \, yr$ after the onset of the mass loss. However, the
mass loss likely slowly increases over time, effectively resetting the
clock and re-creating the shell. The resulting shell may be seen for
up to $10^6\,\rm yr$ following an increase in mass-loss rate.
The precise value depends on ambient density and mass-loss rate.

The limited life time of the ring indicates it cannot pre-date the
present evolutionary phase of the star.  We therefore suggest that
IRAS~02091+6333 presents a case where an AGB star, having undergone
mass loss, perhaps intermittently over some fraction of its AGB life
time, has surrounded itself with a swept-up wall of dust and hydrogen
more than 1\,pc away.

\subsection{Morphology}

Two aspects of the wall need further consideration: first, it is
roughly (but not perfectly) spherical; second, its intensity is
strongest towards and away from the Galactic plane.

The morphology, in the model of the previous subsection, will be
determined by the intrinsic distribution of the stellar wind and by
the structure of the ISM.  The ISM shows structure on many size
scales, however the shell is small compared to its large-scale
structure.  If the shell is running into a region of higher density,
this is most likely to have occured near its outer edge as this is the 
region contributing most to the volume.  This would affect the intensity
distribution, but would be less likely to cause large deviations from
spherical symmetry.

Stellar winds on the AGB tend to be symmetric, until the star
approaches the tip of the AGB (e.g. \citealt{ZCL01}). However, the
effects of the peculiar motion of the star with respect to the ISM
should be considered.  The proper motion indicates a velocity of
45\,km\,s$^{-1}$.  About half of this is due to the effect of the
solar motion and galactic rotation; the remaining velocity vector has
a large uncertainty. The proper motion of the ISM is also not easily
measured, but considerable deviations from circular rotation may occur
in both the young stellar population and the ISM
(e.g. \citealt{ARS84,LZ01}). A recent study of proper-motion vectors
of stars in the CAM OB1 association \citep{Z99}, at a distance of 900
pc and at $\ell \approx 135^{\rm o}$, shows a number of distant
OB-type stars with similar proper motion vectors to
IRAS02091+6333. The relative velocity between star and ISM is
therefore not well determined.  To first order, the effect of stellar
motion is to change the wind velocity with respect to the ISM. The
stalling radius depends on the square root of this velocity, and thus
the shape of the shell is relative insensitive to small peculiar
velocities. But at velocities much larger than 10\,km\,s$^{-1}$, the
effects become large and the star may position itself off-centre over
the life time of the shell.

Supernova remnants often show a bilateral structure very
similar to that of IRAS 02091+633 and these structures are generally
oriented in the same way with respect to the Galactic plane
\citep{Gae98}. The explanation is believed to be shaping by the
Galactic magnetic field, which tends to run along the plane.  The
shaping can either occur directly by the magnetic field, or indirectly
if the ISM is stratified by the field. However, for nebulae as small
as our ring these effects are expected to be negligible. The
similarity is likely accidental. Planetary nebulae do not show a
preferential alignement with respect to the Galactic plane \citep{CAM98}.

\subsection{The Future}

Assuming that the wall around IRAS~02091+6333 is due to AGB--ISM
interaction, the question arises how this structure will evolve.  The
interface between the wind and the ISM will have stalled and the inner
edge of the wall can be considered stationary. The outer edge is
expanding at the velocity of sound, $\sim 1$\,km\,s$^{-1}$.  The
thermal-pulsing AGB lasts several $10^6$\,yr, allowing this edge to
travel several parsec, beyond the phase of observability.  Once the star
enters the post-AGB phase, stellar evolution speeds up dramatically
and the wall will evolve much slower than the inner nebula and its
star.

After the AGB, the circumstellar wind quickly becomes ionized as the
star heats up. At the same time a hot, much faster stellar wind sweeps
up the inner AGB wind, with the swept-up region forming (part of) a
planetary nebula, with a typical radius of 0.1--0.5\,pc \citep{FM94}.
The planetary nebula phase may last $5 \times 10^4$\,yr \citep{ZP91}. 
In our model, this swept-up shell is itself embedded in the much
larger wall. Outer rims and haloes are seen in some planetary nebulae
but they do not have a unique explanation \citep{CSS00} and their
presence cannot be taken as a vindication of this scenario.

In IRAS~02091+6333 the wall is visible as an IRAS shell. Once the
expansion dilutes the wall, this shell becomes less and less distinct.
However, inside the wall is still a region where the density is
much lower, in which the AGB wind keeps the surrounding medium at bay.
Thus, instead of a shell, one can expect the appearance of
an infrared cavity in which the planetary nebula may be located.

Such a cavity has been found around the old nova CK Vul \citep{ELZ02}.
These authors classify CK Vul as a late-thermal-pulse nova, similar
to nova Sakurai. In this case, the nova is expected to be located inside
an old planetary nebula. (A hydrogen nebulosity has been found around
CK Vul but has (sofar) not been  indentified as an old planetary nebula).
The cavity has a radius of approximately 1\,pc. We suggest that
this object may be an example of an old wall.  The presence of other,
very large cavities around planetary nebulae will be the topic of
a future paper.

\section{Summary}

We present the discovery of a large dust shell with a detached
appearance, surrounding a previously unknown but bright AGB star. An
association between the star and the dust shell is likely, given the
statistics of confusion in this region. The spectral type of the star,
magnitude and colour, is used to derive the extinction, and a distance
of $\sim 700\,\rm pc$. The gas mass of the  shell is estimated as
$2.5\, M_\odot$.

The large mass and size of the shell make an origin in a thermal
pulse, as normally assumed for detached shells around AGB stars, unlikely.
Instead we propose a model in which an AGB wind of low mass-loss rate
has swept out a cavity in the ISM; the swept-up material forms the
present shell. We show a typical calculation in which the snowplough
phase lasts for $5 \times 10^5\,$rm. At the end of this phase, the
shell is subsonic, stalls and begins to expand into ISM at its
sound speed. In our model, the shell remains visible for a further
$3 \times 10^5\,$yr. After this, the shell would be visible as an
apparent hole in the ISM surrounding the central AGB star, or the 
later planetary nebula wich follows the AGB evolutionary phase.

A star with a similar size dust shell is R CrB \citep{Gil86}, where a
huge nearly symmetrical dust shell has been discovered. This shell may
have a similar origin. Hydrogen walls are also known around the Sun
and nearby stars, from the interaction between the solar wind and the
ISM (e.g. \citealt{BJ00}): they are a common feature of star--ISM
interactions.

The interaction between an AGB wind and its surrounding ISM has
largely been ignored in the analysis of circumstellar environments,
with the exception of \citet{YPK93}.  The shell of IRAS~02091+6333
suggests that the effects of such an interaction should be further 
explored.

\section*{Acknowledgments}

We are very grateful to Roger Pickard who provided us with the results
of his photometric runs of the central star of the dust shell and to
Maurice Gravin for spectroscopy. Special thanks also go to Sonia
Temporin, Gernot Gr{\"o}mer, Ernst Dorfi and Stefan
Kimeswenger. Fernando Comeron took time to explain to us the various
aspects of low-velocity hydrodynamics.  This research has made use of
the VizieR and SIMBAD databases, maintained by the CDS, of the IPAC
on-line IRAS and MSX databases, and of the IRAS Software Telescope
system maintained by SRON, Groningen.



\begin{thebibliography}{}

\bibitem[\protect\citeauthoryear{{ Lindqvist}, {Olofsson}, {Lucas},
  {Sch\"oier}, {Neri}, {Bujarrabal} \& {Kahane}}{{ Lindqvist}
  et~al.}{1999}]{LOL99}
{ Lindqvist} M.,  {Olofsson} H.,  {Lucas} R.,  {Sch\"oier} F.~L.,  {Neri} R.,
  {Bujarrabal} V.,    {Kahane} C.,  1999, A\&A, 351, L1

\bibitem[\protect\citeauthoryear{{Anantharamaiah}, {Radhakrishnan} \&
  {Shaver}}{{Anantharamaiah} et~al.}{1984}]{ARS84}
{Anantharamaiah} K.~R.,  {Radhakrishnan} V.,    {Shaver} P.~A.,  1984, A\&A,
  138, 131

\bibitem[\protect\citeauthoryear{{Assendorp}, {Bontekoe}, { de Jonge},
  {Kester}, {Roelfsema} \& {Wesselius}}{{Assendorp} et~al.}{1995}]{ABJ95}
{Assendorp} R.,  {Bontekoe} T.~R.,  { de Jonge} A.~R.~W.,  {Kester} D.,
  {Roelfsema} P.~R.,    {Wesselius} P.~R.,  1995, A\&AS, 110, 395

\bibitem[\protect\citeauthoryear{{Ben-Jaffel}, {Puyoo} \&
  {Ratkiewicz}}{{Ben-Jaffel} et~al.}{2000}]{BJ00}
{Ben-Jaffel} L.,  {Puyoo} O.,    {Ratkiewicz} X.,  2000, ApJ, 553, 924

\bibitem[\protect\citeauthoryear{{Boulanger}, {Bernard}, {Lagache} \&
  {Stepnik}}{{Boulanger} et~al.}{2001}]{B01}
{Boulanger} F.,  {Bernard} J.-P.,  {Lagache} G.,    {Stepnik} B.,  2001, in
  Hauser M. H. .~M.,  ed., IAU Symposium 204: The extragalctic infrared
  background and its cosmological implications {Studies of Mira and semiregular
  variables using visual databases}.
p.~47

\bibitem[\protect\citeauthoryear{{Corradi}, {Aznar} \& {Mampaso}}{{Corradi}
  et~al.}{1998}]{CAM98}
{Corradi} R.~L.~M.,  {Aznar} R.,    {Mampaso} A.,  1998, MNRAS, 297, 617

\bibitem[\protect\citeauthoryear{{Corradi}, {Sch{\" o}nberner}, {Steffen} \&
  {Perinotto}}{{Corradi} et~al.}{2000}]{CSS00}
{Corradi} R.~L.~M.,  {Sch{\" o}nberner} D.,  {Steffen} M.,    {Perinotto} M.,
  2000, A\&A, 354, 1071

\bibitem[\protect\citeauthoryear{{de Zeeuw}, R., {de Bruijne}, {Browne} \&
  {Blaauw}}{{de Zeeuw} et~al.}{1999}]{Z99}
{de Zeeuw} T.~P.,  R. H.,  {de Bruijne} J.~H.,  {Browne} A.~G.~A.,    {Blaauw}
  A.,  1999, AJ, 117, 354

\bibitem[\protect\citeauthoryear{{Evans}, {van Loon}, {Zijlstra}, {Pollacco},
  {Smalley}, {Tyne} \& {Eyres}}{{Evans} et~al.}{2002}]{ELZ02}
{Evans} A.,  {van Loon} J.~T.,  {Zijlstra} A.~A.,  {Pollacco} D.,  {Smalley}
  B.,  {Tyne} V.~H.,    {Eyres} S.~P.~S.,  2002, in preparation

\bibitem[\protect\citeauthoryear{{Feast}, {Glass}, {Whitelock} \&
  {Catchpole}}{{Feast} et~al.}{1989}]{FGWC89}
{Feast} M.~W.,  {Glass} I.~S.,  {Whitelock} P.~A.,    {Catchpole} R.~M.,  1989,
  MNRAS, 241, 375

\bibitem[\protect\citeauthoryear{{Flower}}{{Flower}}{1996}]{Flo96}
{Flower} P.~J.,  1996, ApJ, 469, 365

\bibitem[\protect\citeauthoryear{{Frank} \& {Mellema}}{{Frank} \&
  {Mellema}}{1994}]{FM94}
{Frank} A.,  {Mellema} G.,  1994, ApJ, 430, 800

\bibitem[\protect\citeauthoryear{{Gaensler}}{{Gaensler}}{1998}]{Gae98}
{Gaensler} B.~M.,  1998, ApJ, 493, 781+

\bibitem[\protect\citeauthoryear{{Gillet}, {Backman}, {Beichman} \&
  {Neugebauer}}{{Gillet} et~al.}{1986}]{Gil86}
{Gillet} F.~C.,  {Backman} D.~E.,  {Beichman} C.,    {Neugebauer} G.,  1986,
  ApJ, 310, 842

\bibitem[\protect\citeauthoryear{{Glass} \& {Lloyd-Evans}}{{Glass} \&
  {Lloyd-Evans}}{1981}]{GL81}
{Glass} I.,  {Lloyd-Evans} T.,  1981, Nature, 291, 303

\bibitem[\protect\citeauthoryear{{Hashimoto}, {Izumiura}, {Kester} \&
  {Bontekoe}}{{Hashimoto} et~al.}{1998}]{HIKB98}
{Hashimoto} O.,  {Izumiura} H.,  {Kester} D.~J.~M.,    {Bontekoe} T.~R.,  1998,
  A\&A, 329, 213

\bibitem[\protect\citeauthoryear{{Heyer} \& {Tereby}}{{Heyer} \&
  {Tereby}}{1998}]{HT98}
{Heyer} M.~H.,  {Tereby} S.,  1998, ApJ, 502, 265

\bibitem[\protect\citeauthoryear{{Iveciz} \& {Elitzur}}{{Iveciz} \&
  {Elitzur}}{1995}]{IE95}
{Iveciz} Z.,  {Elitzur} M.,  1995, ApJ, 445, 415

\bibitem[\protect\citeauthoryear{{Izumiura}, {Waters}, {de Jong}, {Loup},
  {Bontekoe} \& {Kester}}{{Izumiura} et~al.}{1997}]{IWJ97}
{Izumiura} H.,  {Waters} L.~B.~F.~M.,  {de Jong} T.,  {Loup} C.,  {Bontekoe}
  T.~R.,    {Kester} D.,  1997, A\&A, 323, 499

\bibitem[\protect\citeauthoryear{{Jura}}{{Jura}}{1999}]{J99}
{Jura} M.,  1999, ApJ, 515, 706

\bibitem[\protect\citeauthoryear{{Merch{\' a}n Ben{\' i}tez} \& {Jurado
  Vargas}}{{Merch{\' a}n Ben{\' i}tez} \& {Jurado Vargas}}{2000}]{LZ01}
{Merch{\' a}n Ben{\' i}tez} P.,  {Jurado Vargas} M.,  2000, A\&A, 353, 264

\bibitem[\protect\citeauthoryear{{Olofsson}, {Bergman}, R., {Eriksson},
  {Gustafsson} \& {Bieging}}{{Olofsson} et~al.}{1998}]{OBL98}
{Olofsson} H.,  {Bergman} P.,  R. L.,  {Eriksson} K.,  {Gustafsson} B.,
  {Bieging} J.~H.,  1998, A\&A, 330, L1

\bibitem[\protect\citeauthoryear{{Olofsson}, {Bergman}, R., {Eriksson},
  {Gustafsson} \& {Bieging}}{{Olofsson} et~al.}{2000}]{OBL00}
{Olofsson} H.,  {Bergman} P.,  R. L.,  {Eriksson} K.,  {Gustafsson} B.,
  {Bieging} J.~H.,  2000, A\&A, 353, 583

\bibitem[\protect\citeauthoryear{{Speck}, {Meixner} \& {Knapp}}{{Speck}
  et~al.}{2000}]{SMK00}
{Speck} A.~K.,  {Meixner} M.,    {Knapp} G.~R.,  2000, ApJ, 545, L145

\bibitem[\protect\citeauthoryear{{Spitzer}}{{Spitzer}}{1978}]{Spi78}
{Spitzer} L.,  1978, {Physical Processes in the Interstellar Medium}.
Wiley \&\ Sons

\bibitem[\protect\citeauthoryear{{Tokunaga}}{{Tokunaga}}{2000}]{T00}
{Tokunaga} A.~T.,  2000, in Astrophysical Quantities {}.
p.~143

\bibitem[\protect\citeauthoryear{{Vassiliadis} \& {Wood}}{{Vassiliadis} \&
  {Wood}}{1993}]{VW93}
{Vassiliadis} E.,  {Wood} P.~R.,  1993, ApJ, 413, 641

\bibitem[\protect\citeauthoryear{{Wainscoat}, {Cohen}, {Volk}, {Walker} \&
  {Schwartz}}{{Wainscoat} et~al.}{1992}]{WCV92}
{Wainscoat} R.,  {Cohen} M.,  {Volk} K.,  {Walker} H.~J.,    {Schwartz} D.,
  1992, ApJS, 83, 111

\bibitem[\protect\citeauthoryear{{Weidemann} \& {Sch\"onberner}}{{Weidemann} \&
  {Sch\"onberner}}{1990}]{WS90}
{Weidemann} V.,  {Sch\"onberner} D.,  1990, in From Miras to Planetary Nebulae:
  Which Path for Stellar Evolution? {(Eds. M.O. Mennessier 
and A. Omont) Editions Frontieres, Gif sur Yvette. }
p.~3

\bibitem[\protect\citeauthoryear{{Willems} \& {de Jong}}{{Willems} \& {de
  Jong}}{1986}]{WJ86}
{Willems} F.,  {de Jong} T.,  1986, ApJ, 309, L39

\bibitem[\protect\citeauthoryear{{Wood}}{{Wood}}{1990}]{Woo90}
{Wood} P.~R.,  1990, in From Miras to Planetary Nebulae: Which Path for Stellar
  Evolution? {(Eds. M.O. Mennessier 
and A. Omont) Editions Fontieres, Gif sur Yvette}.
pp 67--84

\bibitem[\protect\citeauthoryear{{Young}, {Phillips} \& {Knapp}}{{Young}
  et~al.}{1993}]{YPK93}
{Young} K.,  {Phillips} T.~G.,    {Knapp} G.~R.,  1993, ApJ, 409, 725

\bibitem[\protect\citeauthoryear{{Zijlstra} \& {Bedding}}{{Zijlstra} \&
  {Bedding}}{2002}]{ZB02}
{Zijlstra} A.~A.,  {Bedding} T.~R.,  2002, MNRAS, accepted for publication

\bibitem[\protect\citeauthoryear{{Zijlstra}, {Chapman}, {te Lintel Hekkert},
  {Likkel}, {Comeron}, {Norris}, {Molster} \& {Cohen}}{{Zijlstra}
  et~al.}{2001}]{ZCL01}
{Zijlstra} A.~A.,  {Chapman} J.~M.,  {te Lintel Hekkert} P.,  {Likkel} L.,
  {Comeron} F.,  {Norris} R.~P.,  {Molster} F.~J.,    {Cohen} R.~J.,  2001,
  MNRAS, 322, 280

\bibitem[\protect\citeauthoryear{{Zijlstra}, {Loup}, {Waters} \& {de
  Jong}}{{Zijlstra} et~al.}{1992}]{ZLWJ92}
{Zijlstra} A.~A.,  {Loup} C.,  {Waters} L.~B.~F.~M.,    {de Jong} T.,  1992,
  A\&A, 265, L5

\bibitem[\protect\citeauthoryear{{Zijlstra} \& {Pottasch}}{{Zijlstra} \&
  {Pottasch}}{1991}]{ZP91}
{Zijlstra} A.~A.,  {Pottasch} S.~R.,  1991, A\&A, 243, 478

\end{thebibliography}

\end{document}